\documentclass[useAMS,usenatbib]{mn2e}
\usepackage{amsbsy}
\usepackage{amssymb}
\usepackage{graphicx}

\usepackage{times}
\usepackage{color}

\newtheorem{definition}{Definition}[section]
\newtheorem{lemma}{Lemma}[section]
\newtheorem{theorem}{Theorem}[section]

\def\Lo{L^{(1)}}
\def\Lt{L^{(2)}}
\def\xio{\boldsymbol{\xi}^{(1)}}
\def\xit{\boldsymbol{\xi}^{(2)}}
\def\xii{\boldsymbol{\xi}^{(i)}}
\def\xin{\xi_0}
\def\xitc{\xi^{(2)}_{c,1}}
\def\xtc{x^{(2)}_{c,1}}
\def\xo{\bmath{x}^{(1)}}
\def\xt{\bmath{x}^{(2)}}
\def\bxi{\bmath{x}^{(i)}}
\def\mxi{x^{(i)}}
\def\mxoeo{x^{(1)}_{E,1}}
\def\mxoet{x^{(1)}_{E,2}}
\def\mxoeth{x^{(1)}_{E,3}}
\def\mxo{x^{(1)}}
\def\mxt{x^{(2)}}
\def\mxii{\xi^{(i)}}
\def\ai{\hat{\boldsymbol{\alpha}}^{(i)}}
\def\a{\boldsymbol{\alpha}}
\def\ao{\hat{\boldsymbol{\alpha}}^{(1)}}
\def\at{\hat{\boldsymbol{\alpha}}^{(2)}}
\def\Do{D_1}
\def\Dt{D_2}
\def\Ds{D_s}
\def\Dot{D_{12}}
\def\Dos{D_{1s}}
\def\Dts{D_{2s}}
\def\mo{m^{(1)}}
\def\mt{m^{(2)}}
\def\Mo{M^{(1)}}
\def\Mt{M^{(2)}}
\def\Mi{M^{(i)}}
\def\Po{\Psi^{(1)}}
\def\Pt{\Psi^{(2)}}
\def\Psii{\Psi^{(i)}}
\def\Si{\Sigma^{(i)}}
\def\si{\sigma^{2(i)}}
\def\So{\Sigma^{(1)}}
\def\Sc{\Sigma_{\mathrm{crit}}}
\def\fo{f^{(1)}}
\def\ft{f^{(2)}}
\def\dfo{\frac{d\fo}{d\mxo}}
\def\dft{\frac{d\ft}{d\mxt}}
\def\J{\det \boldsymbol{\mathsf{J}}}
\def\g{\mathrm{graph}\ }
\def\R{\mathbb{R}_0^+}
\def\to{\theta_{E,1}}
\def\tt{\theta_{E,2}}
\def\tth{\theta_{E,3}}
\def\dres{D_{\mathrm{res}}}

\title{On multiple Einstein rings}

\author[M.C. Werner, J. An and N.W. Evans] {M.C. Werner,$^1$\thanks{E-mail: 
mcw36@ast.cam.ac.uk; nwe@ast.cam.ac.uk} J. An$^{2,3}$\thanks{E-mail: 
jin@dark-cosmology.dk} and N.W. Evans$^1$\footnotemark[1]\\ 
$^1$ Institute of Astronomy, University of Cambridge, 
Madingley Road, Cambridge, CB3 0HA, United Kingdom \\ 
$^2$ Dark Cosmology Centre, Niels Bohr Institute, 
University of Copenhagen, Juliane Maries Vej 30, 2100 Copenhagen~\O, 
Denmark \\ $^3$ Niels Bohr International Academy, Niels 
Bohr Institute, University of Copenhagen, Blegdamsvej 17, 2100 
Copenhagen~\O, Denmark}

\begin{document}

\date{\today}
\pagerange{\pageref{firstpage}--\pageref{lastpage}} \pubyear{0000}
\maketitle

\label{firstpage}

\begin{abstract}
A number of recent surveys for gravitational lenses have found 
examples of double Einstein rings. Here, we investigate analytically 
the occurence of multiple Einstein rings. We prove, under very 
general assumptions, that at most one Einstein ring can arise from a 
mass distribution in a single plane lensing a single background 
source. Two or more Einstein rings can therefore only occur in 
multi-plane lensing. Surprisingly, we show that it is 
possible for a single source to produce more than one Einstein 
ring. If two point masses, or two isothermal spheres, in 
different planes are aligned with observer and source on the 
optical axis, we show that there are up to three Einstein 
rings. We also discuss the image morphologies for these two models 
if axisymmetry is broken, and give the first instances of 
magnification invariants in the case of two lens planes.
\end{abstract}


\section{Introduction}

In his seminal article on the gravitational lensing effect, Einstein
(1936) discussed the circular image of a point-like source, noting
that ``there is no hope of observing this phenomenon directly.'' But,
thanks to advances of instrumentation since then, arcs of partial and
even complete Einstein rings are now being routinely found. Recently,
surveys have found the first instances of multiple Einstein
rings. This includes the partial double Einstein rings of SDSS
J0924+0219 discovered from HST images by the COSMOGRAIL team
\cite{Ei06}, and of SDSS J0946+1006 found by the Sloan Lens ACS Survey
\cite{Ga08}. The Cambridge Sloan Survey of Wide Arcs in the 
Sky (CASSOWARY, Belokurov et al. 2008) has also uncovered a number 
of examples of multiple ring systems, such as CASSOWARY 2. Such
systems may offer valuable insights into the mass distribution of the
lensing galaxies.

From a theoretical point of view, the possibility of forming multiple
Einstein rings has long been known in the case of the strong
deflection limit near the photon sphere of a black hole (for a review,
see e.g. Nemiroff (1993) and references therein). However, a
systematic investigation for the weak deflection limit appears to be
lacking so far, and we present results to this end for the case of one
and two lens planes in this paper. After an outline of the general
lensing setup, the condition for Einstein rings is derived in \S 2. In
the single lens plane case considered in \S 3, we prove that, under rather
general assumptions, multiple Einstein rings cannot arise. We
therefore proceed to two lens planes in \S 4 and consider two singular
isothermal spheres and two point lenses as simple models where
multiple Einstein rings of a single source do, in fact,
occur. In the first example, there are up to two Einstein rings 
due to the singular isothermal spheres in different planes, and 
another ring if the second lens is also luminous. Similarly, two 
point lenses can give rise to up to three Einstein rings overall. 
Therefore, the usual supposition that arcs of multiple Einstein rings
indicate the presence of as many sources at different distances is not
necessarily correct.  We briefly discuss the image configuration in
these cases if axisymmetry is broken. For models with two lens planes,
we also find that analogues of the invariants of the signed
magnification sum (see e.g., Witt \& Mao 1995; Hunter \& Evans 2001) hold in the domains of maximal 
image multiplicity.

Regarding notation, we write $\nabla_{\bmath{v}}$ and
$\Delta_{\bmath{v}}$ for the gradient operator and Laplacian,
respectively, expressed in the same coordinate system as the vector
$\bmath{v}$. Furthermore, $\|\bmath{v}\|$ is the vector norm with
respect to the Euclidean metric, and square brackets $[u]$ denote
functional dependence on the variable $u$. The set
$\R=\{x\in\mathbb R:x\ge0\}$ denotes the set of all non negative reals.
Subscripts can label vector components or 
images according to context. The universal gravitation constant and
the speed of light in vacuum are represented by $G$ and $c$, respectively,
as usual.

\section{Lensing framework}
\subsection{General setup}

\begin{figure}
\centering
\includegraphics[width=1\columnwidth, height=0.5\columnwidth]{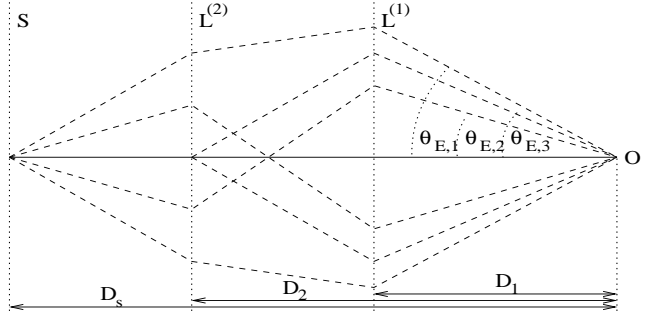}
\caption{Multiple Einstein rings. The source plane $S$ and the two lens planes $\Lo,\Lt$ are at angular diameter distances $\Ds,\Do,\Dt$ from the observer $O$, respectively. A point 
source in $S$ and a luminous lens in $\Lt$ on the optical axis (solid line) a can give rise to three Einstein rings with angular radii $\to,\tt,\tth$ as 
discussed in sections 4.1.2 and 4.2.2. Schematic light rays are shown as dashed lines.}
\label{fig:einsteinrings}
\end{figure}

Gravitational lensing in the weak deflection limit is conveniently
described in terms of the impulse approximation, with piecewise
straight light rays in flat space between the source, lens planes and
observer (for a comprehensive introduction see e.g. Schneider,
Ehlers \& Falco 1999). We consider a point source in the source plane
$S$ and two lens planes $\Lo,\ \Lt$ with angular diameter distances
$\Do \leq \Dt \leq \Ds$ between the observer and $\Lo,\ \Lt,\ S$, also
$\Dot,\Dos$ between $\Lo$ and $\Lt, S$, and $\Dts$ between $\Lt, S$,
respectively. This setup is illustrated in figure \ref{fig:einsteinrings}. 
Using Cartesian coordinates $\boldsymbol{\eta}$ in $S$
and $\xio,\ \xit$ in $\Lo,\ \Lt$ measured from some optical axis, and
$\ao,\at$ for the deflection angles of the light rays crossing the
respective lens planes, the lens equations become,
\begin{eqnarray}
\xit&=&\frac{\Dt}{\Do} \xio-\Dot\ao [\xio], \label{lenseq1}\\
\boldsymbol{\eta}&=&\frac{\Ds}{\Do} \xio-\Dos\ao[\xio]-\Dts\at[\xit].
\label{lenseq2}
\end{eqnarray}
Assuming that the optical axis passes through the centre of the
projected mass distribution in $\Lo$, the deflection potentials in
$\Lo,\Lt$ can be written with an overall scaling factor proportional
to the lens mass such that
\begin{eqnarray}
\Po[\xio]&=&\Mo \fo[\xio], \label{deflect1}\\
\Pt[\xit]&=&\Mt \ft[\xit-\xit_c],
\label{deflect2}
\end{eqnarray}
where the centre of the projected mass distribution in $\Lt$ may be
taken as $\xit_c=(\xitc,0)$ without loss of generality. Then the
deflection angles are
\begin{equation}
\ai[\xii]=\nabla_{\xii}\Psii[\xii], \ i \in \{1,2\}.
\label{angle}
\end{equation}
In this notation, the deflection potentials are solutions of Poisson's
equation, corresponding to Einstein's equation in this quasi-Newtonian 
approximation,
\begin{equation}
\Delta_{\xii}\Psii[\xii]=\frac{8\pi G}{c^2}\Si[\xii], \ i \in \{1,2\},
\label{poisson}
\end{equation}
where $\Si$ is the projected surface mass density in the $i$th lens
plane. Given a model for the surface density $\Si$, the integration
constants for the corresponding deflection potential $\Psii$ are
chosen such that the norm of the deflection angle at infinity is as
small as possible, that is, zero for realistic lens models.

While $\xio/\Do$ would correspond to the usual angular coordinate in
$\Lo$, we shall find the space of parameters more convenient if the
following normalization is used \citep{Er93},
\begin{equation}
\xo=\frac{\xio}{\xin},
\ \xt=\frac{\xit/\Dt}{\xin/\Do},
\ \bmath{y}=\frac{\boldeta/\Ds}{\xin/\Do},
\label{norm1}
\end{equation}
with parameters
\begin{equation}
\xin=
\Do\left(\frac{\mu^{(1)}+\mu^{(2)}}{\Ds}\right)^{1/2},
\label{norm2}
\end{equation}
where $\mu^{(1)}=\Mo(\Dos/\Do)$, and
$\mu^{(2)}=\Mt(\Dts/\Dt)$ and
\begin{equation}
\beta=\frac{\Dot\Ds}{\Dos\Dt}=1-\frac{\Dts\Do}{\Dos\Dt}
\label{beta}
\end{equation}
Here, $0\le\beta\le1$. In particular, $\beta=0$ if and only if $\Dot=0$ and
$\beta=1$ if and only if $\Dts=0$, provided that $\Do\ne0$, $\Dt\ne0$ and $\Ds\ne0$.
Then, the lens equations (\ref{lenseq1}), (\ref{lenseq2}) can be rewritten thus
\begin{eqnarray}
\xt &=& \xo-\beta \mo\bmath\nabla_{\xo} \fo[\xo], \label{lenseq3}\\
\bmath{y} &=& \xo-\mo\bmath\nabla_{\xo} \fo[\xo]
\nonumber\\
&-&\mt\bmath\nabla_{\xt}\ft[\xt-\xt_{\mathrm c}],
\label{lenseq4}
\end{eqnarray}
introducing the normalized mass parameters
%
\begin{eqnarray}
\mo&=&\frac{\mu^{(1)}}{\mu^{(1)}+\mu^{(2)}}, \label{norm3}\\
\mt&=&\frac{\mu^{(2)}}{\mu^{(1)}+\mu^{(2)}},
\label{norm4}
\end{eqnarray}
such that $\mo+\mt=1$.

\subsection{Einstein rings}

Einstein rings, within the geometrical optics approximation of the
standard lensing framework, are infinitely magnified, circularly
symmetric images. We therefore stipulate circular symmetry about the
optical axis such that $\xtc=0$ and
\[
f^{(i)}[\bxi]=f^{i}[\|\bxi\|], \ i \in \{1,2\}, \ \mbox{and} \ \|\bxi \|\equiv\mxi
\]
is used for notational simplicity. But then the lens equation (\ref{lenseq3}) implies that
$\mxt=\mxt[\mxo]$ and (\ref{lenseq4}) becomes
\begin{eqnarray}
\bmath{y}&=&\xo-\xo\frac{\mo}{\mxo}\dfo \nonumber \\
&-&\xo\frac{\mt}{\mxt[\mxo]}\left(1-\beta\mo\dfo\right)\dft[\mxo]\nonumber\\
&\equiv&\xo F[\mxo].
\label{lenseq5}
\end{eqnarray}
Introducing plane polar coordinates $(\mxo, \phi)$, the Jacobian
determinant of the lensing map (\ref{lenseq5}) is therefore
\begin{equation}
\J=\frac{1}{\mxo}
\left| \begin{array}{cc}
\frac{\partial y_1}{\partial \mxo} & \frac{\partial y_1}{\partial \phi} \\ \\
\frac{\partial y_2}{\partial \mxo} & \frac{\partial y_2}{\partial \phi}
\end{array} \right|
=F\left(F+\mxo \frac{dF}{d\mxo}\right).
\label{jacob}
\end{equation}
The condition for critical curves $\J=0$ gives rise to two classes of
critical curves. The first, given by $F[\mxo]=0$, defines tangential
critical circles in $\Lo$ which are solutions of the lens equation
(\ref{lenseq5}), and hence infinitely magnified images, and map to the
caustic point $\bmath{y}=\bmath{0}$. The other solution $F+\mxo
dF/d\mxo=0$ gives radial critical circles in $\Lo$ which map to
caustic circles and hence define domains of constant image
multiplicity in $S$ \cite[ p. 233]{Sc99}.
\begin{definition}
An Einstein ring is a circular, critical image of a point source at $\bmath{y}=\bmath{0}$ whose radius is a solution of $F[\mxo]=0, \ \mxo>0$.
\label{def1}
\end{definition}
Finally, we note that if $\bmath{y}\neq \bmath{0}$ so that the
axisymmetry is broken, discrete images are formed at some $\xo$ which
have finite signed magnification \cite[ p. 162]{Sc99}
\begin{equation}
\mu[\xo]=\frac{1}{\J[\xo]}.
\label{jacob2}
\end{equation}

\section{One lens plane}

We continue by specializing the previous discussion to the simpler
case of a single lens plane $\Lo=\Lt$ such that $\mt=0,
\ \beta=0$. Notice, then, that (\ref{angle}), (\ref{norm1}) and
(\ref{norm2}) yield
\[
\nabla_{\xo}\fo=\frac{\Do\Dos\Mo}{\Ds\xin}\nabla_{\xio}\fo=\frac{\Do\Dos}{\Ds\xin}\ao\equiv\a,
\]
which is the normalized deflection angle in the standard form \cite[
  p. 158]{Sc99}. Similarly, the Poisson equation (\ref{poisson})
  becomes
\[
\Mo\Delta_{\xio}\fo=\frac{8\pi G}{c^2}\So
\Rightarrow \Delta_{\xo}\fo=2\frac{\So}{\Sc}\equiv 2\kappa
\]
with the usual definition of the critical surface density
$\Sc=c^2\Ds/(4\pi G \Do\Dos)$. Using definition \ref{def1} and writing
$\|\a \|\equiv \alpha=d\fo/d\mxo$, the problem of finding Einstein
rings therefore reduces to a fixed point equation
\begin{equation}
\alpha[\mxo] = \mxo, \ \mxo>0,
\label{fix}
\end{equation}
subject to the Poisson equation for the given mass distribution of the lens,
\begin{equation}
\Delta_{\xo} \fo[\mxo] = 2\kappa[\mxo].
\label{poisson2}
\end{equation}
Hence, we need to define general, yet astrophysically sensible lens
models $\kappa[\mxo]$ that allow for Einstein rings. Apart from the
circular symmetry inherent in the problem, it would be plausible to
stipulate that $\kappa$ decreases monotonically with $\mxo$. However,
we shall use a condition even weaker than monotonicity, namely that,
at every radius, the surface density be smaller than the average
density of the mass enclosed. This is natural for self-gravitating and
hence centrally condensed systems, and a more specific model for the
mass distribution need not be assumed.
\begin{definition}
The gravitational lens is defined by a normalized surface density
$\kappa$ as mass model such that the following conditions are
fulfilled.
\begin{enumerate}
\item{Continuity: $\kappa: \Lo \rightarrow \R$ be a continuous function except at $\bmath{0}\in \Lo$ for singular lenses.}
\item{Circular symmetry: $\kappa=\kappa[\mxo]$ only.}
\item{Finiteness: $\kappa < \infty$ for $\mxo>0$, $\kappa[0]=1/C_1$
  with constant $C_1 \geq 0$, and $\lim_{\mxo\rightarrow
    \infty}\kappa\mxo=C_2$ with constant $0\leq C_2 < \infty$.}
\item{Self-gravitation: $\kappa[\mxo]<\bar{\kappa}[\mxo]$ where
  $\bar{\kappa}[\mxo]=\frac{2}{(\mxo)^2} \int_0^{\mxo}\kappa[x]xdx$.}
\end{enumerate}
\label{def2}
\end{definition}
For example, the point lens has $C_1=0, \ C_2=0$, and the isothermal
sphere $C_1=0, \ C_2>0$ in our notation. Both are usually regarded as
singular, since their projected surface densities diverge at the
centre. Moreover, the total mass is infinite in the isothermal case,
and we should expect $C_2=0$ for more realistic lenses. However, it
turns out that we do not need to require this for our purposes and
hence define singular lenses simply as follows.
\begin{definition}
A gravitational lens is called singular if, and only if, $C_1=0$, and is called non-singular otherwise.
\label{def3}
\end{definition}
With these definitions, one can now prove that there is at most one
Einstein ring. This result follows from existence, whose necessary and
sufficient condition is established in theorem \ref{existence}, and
uniqueness, shown in theorem \ref{uniqueness} below. First of all,
however, we shall need a lemma regarding the surface density, and a
lemma concerning the deflection angle.
\begin{lemma}
Given a gravitational lens in the sense of definitions \ref{def2} and
\ref{def3}, then the normalized surface density fulfils
$\kappa[\mxo]<\kappa[0] \ \forall \ \mxo>0$.
\begin{proof}
If the lens is singular, then the lemma follows immediately from
definition \ref{def2}(iii). Otherwise, by definition \ref{def2}(iv),
we have
\[
\kappa[\mxo] < \bar{\kappa}[\mxo] < \max_{0\leq x \leq \mxo}\kappa[x] \ \forall \ \mxo. 
\]
Suppose that this maximum is attained at some $x' <\mxo$. Then the
previous inequalities hold for $\kappa[x']$ as well, with maximum at
$x''<x'$, say. Repeating this argument sufficiently often shows that
$\kappa[\mxo]<\kappa[0]$, as required.
\end{proof}
\label{lemma1}
\end{lemma}

\begin{lemma}
Given a gravitational lens in the sense of definitions \ref{def2} and
\ref{def3}, then the normalized deflection angle $\alpha$ has the
following properties.
\begin{enumerate}
\item{Smoothness: $\alpha: \Lo \rightarrow \R$ is a smooth function except at $\bmath{0}\in \Lo$ for singular lenses.}
\item{Circular symmetry: $\alpha=\alpha[\mxo]$ only.}
\item{Properties at the centre: For non-singular lenses,
  $\alpha[0]=0$ and the derivative $d\alpha/d\mxo[0]=\kappa[0]$, for
  singular lenses $\alpha[0]>0$.}
\item{Finiteness: $\lim_{\mxo\rightarrow \infty}\alpha=A$ with constant $A < \infty$.}
\end{enumerate}
\begin{proof}
Definition \ref{def2}(i) implies property (i), and property (ii)
follows immediately from definition \ref{def2}(ii) and the uniqueness
of solutions of Poisson's equation. Circular symmetry and smoothness at the centre 
imply that $\alpha[0]=0$ for non-singular lenses, and $\alpha[0]$ is some positive, possibly infinite, value for singular lenses. 
Now integrate Poisson's equation (\ref{poisson2}) to find
\begin{eqnarray}
2\kappa&=&\frac{1}{\mxo}\frac{d}{d\mxo}(\alpha \mxo)\qquad \Rightarrow \label{angle2}\\
\alpha[\mxo]&=&\frac{2}{\mxo}\int_0^{\mxo}\kappa[x]x dx. \label{angle3}
\end{eqnarray}
using $\alpha[0]=0$. Then (\ref{angle2}) implies the second property of non-singular lenses in (iii),
\begin{eqnarray*}
\kappa[0]&=&\frac{1}{2}\lim_{\mxo\rightarrow
  0}\left(\frac{d\alpha}{d\mxo}+\frac{\alpha
  +\frac{d\alpha}{d\mxo}\mxo+\mathcal{O}[(\mxo)^2]}{\mxo}\right)\\
&=&\frac{d\alpha}{d\mxo}[0]
\end{eqnarray*}
using again $\alpha[0]=0$, and property (iv) follows from equation (\ref{angle3}) and definition \ref{def2}(iii) 
since
\[
\lim_{\mxo\rightarrow \infty}\alpha=2\lim_{\mxo\rightarrow \infty}\kappa \mxo=2C_2,
\]
so $A=2C_2 < \infty$, as required. This is also true for singular lenses 
where the integration of (\ref{angle2}) has to start at some $x>0$, so that the limit for $\alpha$ as $\mxo\rightarrow \infty$ is 
the same as before plus some finite integration constant.
\end{proof}
\label{lemma2}
\end{lemma}

\begin{theorem}[Existence]
Given a gravitational lens in the sense of definitions \ref{def2} and
\ref{def3}, then at least one Einstein ring exists if, and only if, the 
condition $\kappa[0]>1$ holds.
\begin{proof}
First, we prove the sufficiency of the condition: $\kappa[0]>1 \Rightarrow \ \exists$
Einstein ring. Let us consider the graph of the deflection angle
\[
\g\alpha=(\mxo,\alpha(\mxo))\in \R\times\R.
\]
Then, according to equation (\ref{fix}), the existence of an Einstein
ring is equivalent to the existence of a fixed point and hence an
intersection of $\g \alpha$ with the diagonal $(\mxo,\mxo) \in
\R\times\R$ at some $\mxo>0$. Let $A_1=\{(x,y)\in \R\times\R: y >
\mxo\}$ and $A_2=\{(x,y)\in \R\times\R: y < \mxo\}$ be two domains to
the left and right of the diagonal, respectively. For non-singular
lenses, lemma \ref{lemma2}(iii) implies that $\mxo=0$ is a fixed point
but no Einstein ring, and $\kappa[0]>1 \Rightarrow d\alpha/d\mxo[0]>1
\Rightarrow \g \alpha \in A_1$ for $\mxo\rightarrow 0$. This is also
true for singular lenses because here $\alpha[0]>0$. On the other
hand, lemma \ref{lemma2}(iv) implies that $\g \alpha \in A_2$ for
$\mxo \rightarrow \infty$. This ensures the existence of at least one
fixed point with $\mxo>0$ and hence Einstein ring.

Now we show necessity: $\exists$ Einstein ring $\Rightarrow
\ \kappa[0]>1$. This statement is equivalent to its contraposition,
$\kappa[0]\leq 1 \Rightarrow \ \nexists$ Einstein ring, which we prove
by contradiction. So suppose $\kappa[0]\leq 1$ and $\exists$ Einstein
ring, then there is some $\mxo_E>0$ such that $\alpha[\mxo_E]=\mxo_E$
by equation (\ref{fix}). By recasting Poisson's equation
(\ref{poisson2}) and using lemma \ref{lemma1},
\begin{eqnarray*}
\frac{d\alpha}{d\mxo}[\mxo_E]&=&2\kappa[\mxo_E]-1< 2\kappa[0]-1 \qquad \Rightarrow \\
\frac{d\alpha}{d\mxo}[\mxo_E]&<&1,
\end{eqnarray*}
since by assumption $\kappa[0]\leq 1$. But on the other hand, for a
fixed point $\mxo_E>0$ to exist, we must require
$d\alpha/d\mxo[\mxo_E]\geq 1$. To see this, notice first of all that
the assumption $\kappa[0]\leq 1$ also means, by definitions
\ref{def2}(iii) and \ref{def3}, that we only need to consider
non-singular lenses here. Now, by lemma \ref{lemma2}(iii), this
implies that $d\alpha/d\mxo[0]\leq 1$. By differentiating
(\ref{angle2}) and using l'H\^{o}pital's rule,
\[
\frac{d^2\alpha}{dx^{(1)2}}[0]=\frac{4}{3}\frac{d\kappa}{d\mxo}[0]<0 \ \mbox{by lemma \ref{lemma1},}
\]
 so even if $\kappa[0]=1, \ \g \alpha \in A_2$ for small $\mxo$, and
 the result follows. This completes the proof by contradiction and
 hence the proof of the theorem.
\end{proof}
\label{existence}
\end{theorem}

\begin{theorem}[Uniqueness]
Given a gravitational lens in the sense of definitions \ref{def2} and
\ref{def3}, then if an Einstein ring exists, there is exactly one.
\begin{proof}
The existence of an Einstein ring $E$ implies that $\kappa[0]>1$ by
theorem \ref{existence}, and that there is some $\mxo_E>0$ such that
$\alpha[\mxo_E]=\mxo_E$ by equation (\ref{fix}). Using the integral
equation (\ref{angle3}) and definition \ref{def2}(iv), we obtain
\[
1=\frac{\alpha[\mxo_E]}{\mxo_E}=\frac{2}{(\mxo_E)^2}\int_0^{\mxo_E}\kappa[x]xdx=\bar{\kappa}[\mxo_E].
\]
Hence by the Poisson equation (\ref{angle2}) and definition \ref{def2}(iv),
\begin{eqnarray*}
\frac{d\alpha}{d\mxo}[\mxo_E]&=&2\kappa[\mxo_E]-1<2\bar{\kappa}[\mxo_E]-1 \qquad \Rightarrow \\
\frac{d\alpha}{d\mxo}[\mxo_E]&<&1.
\label{angle4}
\end{eqnarray*}
there is another Einstein ring $E'$ at $\mxo_{E'}>\mxo_E$. But by the
same token used in the sufficiency proof of the previous theorem, $\g
\alpha \in A_1$ for small $\mxo$ and $\g \alpha \in A_2$ for
$\mxo\rightarrow \infty$, so we need $d\alpha/d\mxo[\mxo_{E'}]\geq 1$
for the fixed point of $E'$ to exist. This cannot be according to
(\ref{angle4}), and the result follows.
\end{proof}
\label{uniqueness}
\end{theorem}
Since multiple Einstein rings cannot occur in a one lens plane setting
by theorems \ref{existence} and \ref{uniqueness}, we now discuss to
two simple models in the two lens plane case.

\section{Two lens planes}

\subsection{Singular isothermal spheres}
\subsubsection{Lens equation}
The first system of two lenses discussed here consists of a singular
isothermal sphere both in $\Lo$ and $\Lt$. This is hence a special
case of the cored, spherically symmetric lenses in two lens planes
considered by Kochanek \& Apostolakis (1988) in the context of a
numerical study of lensing cross-sections. However, one can easily
extend this model, which has two identical lenses, to include the
effect of different surface densities. Because of the circular
symmetry required for Einstein rings, we shall assume the two singular
isothermal spheres to be centred on the optical axis so that $\xitc=0$
again, and consider projected surface densities \cite[ p. 243]{Sc99}
\[
\Si[\xii]=\frac{\si}{2G\mxii}, \ i \in \{1,2\},
\] 
where $\mxii\equiv \|\xii \|$ and $\si$ is line of sight velocity
dispersion of the isothermal sphere in the $i$th lens plane. Now
according to Poisson's equation (\ref{poisson}), the deflection
potentials (\ref{deflect1}), (\ref{deflect2}) become
\[
\Psii[\xii]=\Mi\mxii \ \mbox{where}\ \Mi=\frac{4\pi\si}{c^2},\ i\in \{1,2\},
\]
and hence the lens equation (\ref{lenseq5}),
\begin{equation}
\bmath{y}=\xo\left(1-\frac{m_\pm}{\mxo}\right)
\label{lenseq6}
\end{equation}
in which $\mo,\mt$ from equations (\ref{norm3}) and (\ref{norm4}) have
been combined to define a new mass parameter in this case,
\begin{equation}
m_\pm=\xin\left(\mo\pm\frac{\Dt}{\Do}\mt\right)
\label{norm5}
\end{equation}
where the positive sign is valid for $\mxo\geq \beta\xin \mo$ and the
negative sign for $\mxo < \beta \xin\mo$.

\subsubsection{Einstein rings}
Now if $\bmath{y}=\bmath{0},\ F[\mxo]=1-m_\pm/\mxo$, no radial
critical circles exist. Definition \ref{def1} and (\ref{lenseq6})
imply that the radii of Einstein rings are given by
\begin{equation}
\mxo=m_\pm,
\label{iso}
\end{equation}
subject to the two domains of the mass parameter (\ref{norm5}). It
turns out, then, that the point source in $S$ lensed by both
isothermal spheres in $\Lo$ and $\Lt$ always produces one Einstein
ring of radius
\begin{equation}
\mxoeo=\frac1{\Ds^{1/2}}\frac{\Mo\Dos+\Mt\Dts}{\left(\mu^{(1)}+\mu^{(2)}\right)^{1/2}},
\end{equation}
and, in addition, provided $\Mo/\Mt < \Dt/\Do$, another one of radius
\begin{equation}
\mxoet=\frac1{\Ds^{1/2}}\frac{\Mo\Dos-\Mt\Dts}{\left(\mu^{(1)}+\mu^{(2)}\right)^{1/2}}.
\end{equation}
Furthermore, there is a third Einstein ring of the isothermal sphere in $\Lt$
lensed by the one in $\Lo$, assuming, of course, that at least the
former is luminous. Because the lens in $\Lt$ is extended, this
second Einstein ring is not a circle but has some radial width. We
therefore define the radius of this third Einstein ring $\mxoeth$ in
$\Lo$ to be that of the circular image produced by the centre of the 
lens (the most luminous part) in $\Lt$. Its radius can be
read off immediately from the previous results by letting $\Mt=0$ and
replacing $\Ds,\Dos$ by $\Dt,\Dot$, respectively, so that there are
two different definitions of $\xin$, one for the two rings due to the point source, and one for the ring due to the 
isothermal sphere in $\Lt$. Hence, using
(\ref{norm2}) and (\ref{norm5}),
\[
\mxoeth=\left(\frac{\Do\Dot}{\Dt}\Mo\right)^{1/2}
\]
In view of observational applications, it is more convenient to
express these Einstein ring radii in terms of angular coordinates in
$\Lo$. Given the small angles approximation, one can take the angular
radius to be $\theta=\xi^{(1)}/\Do=\mxo\xin/\Do$ using
(\ref{norm1}). Hence for the three Einstein rings,
\begin{eqnarray}
\to&=&\frac{\Mo\Dos+\Mt\Dts}{\Ds}
\label{rings}\\
\tt&=&\frac{\Mo\Dos-\Mt\Dts}{\Ds}
\ \ \mbox{if}\ \ \frac{\Mo}{\Mt}<\frac{\Dt}{\Do},\label{rings2}\\
\tth&=&\frac{\Dot}{\Dt}\Mo.\label{rings3}
\end{eqnarray}

\subsubsection{Image configuration}
According to the lens equation (\ref{lenseq6}), this system of two
singular isothermal spheres at different distances is a modification
of the well-known single lens plane case in the sense that the two
instances $m_\pm$ have to be distinguished here.  It turns out, then,
that up to four images can be obtained in the present case. To be more
precise, consider the image configurations if axisymmetry is broken
such that $\bmath{y}\neq\bmath{0}$.  Without loss of generality, one
may take $y_1\equiv y>0, y_2=0$ so that $\bmath{y}=(y,0)$. Images are
hence collinear with the centre of $\Lo$ and the source such that
$\xo=(x,0)$, say. Therefore (\ref{lenseq6}) and (\ref{norm5}) imply
the following image positions $x_i, \ i \in \{1, \ldots, 4\},$ for
given domains of $\bmath{y}\in S$,
\begin{eqnarray}
x_1&=&y+m_{+}>0 \ \ \mbox{for} \label{iso1} \\
y&\geq& -m_{+}+\beta\xin\mo = -\frac{\xin}{\Do}\frac{\Dts}{\Dt}\frac{\Mo\Do+\Mt\Dt}{\mu^{(1)}+\mu^{(2)}},
\nonumber \\
x_2&=&y+m_{-}>0 \ \ \mbox{for} \label{iso2} \\
y &<& -m_{-}+\beta \xin \mo = -\frac{\xin}{\Do}\frac{\Dts}{\Dt}\frac{\Mo\Do-\Mt\Dt}{\mu^{(1)}+\mu^{(2)}},
\nonumber \\
x_3&=&y-m_{+}<0 \ \ \mbox{for} \label{iso3} \\
y &\leq& m_{+}-\beta\xin\mo = \frac{\xin}{\Do}\frac{\Dts}{\Dt}\frac{\Mo\Do+\Mt\Dt}{\mu^{(1)}+\mu^{(2)}},
\nonumber \\
x_4&=&y-m_{-}<0 \ \ \mbox{for} \label{iso4} \\
y &>& m_{-}-\beta \xin \mo = \frac{\xin}{\Do}\frac{\Dts}{\Dt}\frac{\Mo\Do-\Mt\Dt}{\mu^{(1)}+\mu^{(2)}}.
\nonumber
\end{eqnarray}
The condition on $y$ for (\ref{iso1}) to hold is clearly always
fulfilled, and the condition for (\ref{iso3}) defines a cut circle in
$S$ within which another image occurs. Hence the images given by $x_1$
and $x_3$ correspond to the single lens plane case if $\beta=0$ (see
e.g. Schneider et al. 1999, p. 244).  Similarly, the condition on $y$
in (\ref{iso2}) defines a second cut circle. But because we need to
keep $y$ positive by setup, this cut circle only exists if $\Mo<\Mt
\Dt/\Do$, the same condition as for the second Einstein ring
(\ref{rings2}) discussed in the previous section. Furthermore, if the
source is outside of the second cut circle, then one image $x_4$ now
occurs on the opposite (i.e., negative) side of $\Lo$. If the source
is inside the second cut circle, another image $x_2$ appears on the
same side of $\Lo$ as the source. Finally, if $y\rightarrow 0$, then
all image positions approximate to the Einstein rings given by
(\ref{iso}), as expected. However, it is interesting that, unlike in
the single lens plane case, the cut circles and the Einstein rings do
not coincide here.

Now assume that $y$ is in a maximal domain in $S$ such that all four
possible images $x_i$ (\ref{iso1}-\ref{iso4}) are present. Then, given
that $\J=1-m_\pm/\mxo$ by (\ref{jacob}), we can directly evaluate the
sum of the signed magnifications (\ref{jacob2}) of the images to find
\begin{equation}
\sum_{i=1}^4 \mu[x_i]=4.
\label{isomag}
\end{equation}
This is therefore a magnification invariant in the two lens plane case.

\subsection{Point lenses}
\subsubsection{Lens equation}
We now turn to two point lenses. A thorough study of the caustic
structure for two point lenses in general positions within
three-dimensional space was done by Erdl \& Schneider (1993). However,
the aspect of multiple Einstein rings was not studied there
explicitly, so we present results to this end below. Using a setup
analogous to the one in the previous section, consider two point
masses $M_1,M_2$ on the optical axis in $\Lo,\Lt$, respectively. In
this case, then, the surface densities are given by \cite[
p. 239]{Sc99}
\[
\Si[\xii]=M_i \delta^{2(i)}[\xii], \ i \in \{1,2\},
\]
where $\delta^{2(i)}$ denotes the two-dimensional delta function of
the $i$th lens plane. Hence, the corresponding deflection potentials
\[
\Psii[\xii]=\Mi \ln\left[\frac{\mxii}{\xin}\right] \ \mbox{where}\ \Mi=\frac{4GM_i}{c^2},\ i \in \{1,2\},
\]
follow from (\ref{poisson}), and the lens equation (\ref{lenseq5}) becomes
\begin{equation}
\bmath{y}=\xo\left(1-\frac{\mo}{(\mxo)^2}-\frac{\mt}{(\mxo)^2-\beta\mo}\right)
\label{lenseq7}
\end{equation}
using (\ref{beta}), (\ref{norm3}) and (\ref{norm4}) \cite{Er93}. 
Recall also that the range of the parameters used in
(\ref{lenseq7}) is $0\le\beta,\mo,\mt \le1$ by definition.

\subsubsection{Einstein rings} 
Again, according to definition \ref{def1}, Einstein rings are given by
positive solutions of $F(\mxo)=0$ for $\bmath{y}=\bmath{0}$, which
is a quartic in $\mxo$ by lens equation (\ref{lenseq7}). Hence,
letting $(\mxo)^2\equiv r$, we seek solutions
\[
F(r)=1-\frac{\mo}{r}-\frac{\mt}{r-\beta\mo}=0, \ r>0,
\]
and obtain the squares of Einstein ring radii
\begin{equation}
r_E=\frac{1}{2}\left(1+\beta\mo \pm \sqrt{(1+\beta\mo)^2 -4\beta(\mo)^2}\right)
\label{rings4}
\end{equation}
using (\ref{norm3}) and (\ref{norm4}). Notice that
\begin{eqnarray}
(1+\beta\mo)^2&>&4\beta(\mo)^2 \label{ineq}\\
\mbox{unless} \ \beta= 1 &\land& \mo= 1, \ \mbox{and}\nonumber \\
1+\beta\mo&>&\sqrt{(1+\beta\mo)^2-4\beta(\mo)^2} \nonumber \\ 
\mbox{unless}\ \beta= 0 &\lor& \mo= 0.\nonumber
\end{eqnarray}
Therefore, in the general case $0<\beta<1, \ 0<\mo<1$, there are
exactly two Einstein rings with radii $\mxoeo,\mxoet$ from
(\ref{rings4}) due to a point source in $S$ and the point lenses in
$\Lo$ and $\Lt$. Otherwise, one ring becomes a critical point on the
optical axis, and we recover the case of a single Einstein ring, as
expected.

If, in addition, the point lens in $\Lt$ is also luminous, it produces
a third Einstein ring at radius $\mxoeth$ due to the point lens in
$\Lo$. As before, this radius can be obtained simply by setting
$\mt=0$ and identifying $\Ds,\Dos$ with $\Dt, \Dot$, respectively. The
corresponding angular radii are again given by
$\theta=\xi^{(1)}/\Do=\mxo\xin/\Do$, and we find
\begin{eqnarray}
\to&=&\sqrt{\frac{\Dos}{\Do\Ds}\left(p+\sqrt{p^2-\beta(\Mo)^2}\right)}, \label{rings5}\\
\tt&=&\sqrt{\frac{\Dos}{\Do\Ds}\left(p-\sqrt{p^2-\beta(\Mo)^2}\right)} , \label{rings6}\\
\tth&=&\sqrt{\frac{\Dot}{\Do\Dt}\Mo} \label{rings7}, \ \mbox{with mass parameter}\\
p&\equiv&\frac{1+\beta}{2}\Mo+\frac{1-\beta}{2}\Mt\nonumber
\end{eqnarray}
using (\ref{norm2}), (\ref{beta}), (\ref{norm3}) and (\ref{norm4}).

\subsubsection{Image configuration}
Now consider the case when $\bmath{y}\neq \bmath{0}$. As before, we
can choose without loss of generality $\bmath{y}=(y,0), \ y>0$. The
images are collinear with the source so that $\xo=(x,0)$. Hence
solutions of the lens equation (\ref{lenseq7}) are defined by a
quintic in $x$,
\[
0=x^5-yx^4-(1+\beta\mo)x^3+y\beta\mo x^2+\beta(\mo)^2 x.
\]
We immediately have a solution on the optical axis, $x_5=0$, but this
image is infinitely demagnified, and therefore invisible, since its
signed magnification is given by
\[
\mu[x_5]=\frac{1}{\J [x_5]}=0 \ \mbox{because}\ \lim_{x\rightarrow x_5}F[(x,0)]=-\infty
\]
using (\ref{jacob2}). The {remaining roots $x_i, \ i \in \{1, \ldots, 
4\},$ are solutions of the quartic equation
\begin{equation}
0=x^4-yx^3-(1+\beta\mo)x^2+y\beta\mo x+\beta(\mo)^2.
\label{lenseq8}
\end{equation}
We note that this immediately implies that the roots satisfy
\[
\sum_{i=1}^4 x_i = y.
\]
Now Descartes' rule of signs shows that (\ref{lenseq8}) has either
two positive and two negative real roots, two positive real roots and
a complex conjugate pair, two negative real roots and a complex
conjugate pair, or two complex conjugate pairs of roots. So in the
first case, there are four images, in the second and third case, two
images each, and in the last case, none. As usual, these cases can be
distinguished by means of the cubic resolvent of (\ref{lenseq8}). With
the standard substitution $z\equiv x-y/4$ we can eliminate the cubic
term in (\ref{lenseq8}) to obtain a reduced quartic in $z$,
\begin{eqnarray*}
0&=&z^4+Pz^2+Qz+R, \ \mbox{where} \\
P&=&-\frac{3y^2}{8}-(1+\beta\mo),\\
Q&=&-\frac{y^3}{8}-\frac{y}{2}(1-\beta\mo),\\
R&=&-\frac{3y^4}{256}-\frac{y^2}{16}(1-3\beta\mo)+\beta(\mo)^2.
\end{eqnarray*}
Hence the resolvent of (\ref{lenseq8}) is given by the cubic equation \cite[ p. 121]{Br85}
\begin{equation}
0=z^3+Az^2+Bz+C
\label{cubic}
\end{equation}
where
\[
A=2P, \ B=P^2-4R, \ C=-Q^2.
\]
Notice, then, that $A<0$ and
\[
B=\frac{3y^4}{16}+y^2+(1+\beta\mo)^2-4\beta(\mo)^2>0
\]
using (\ref{ineq}). Applying Descartes' rule of signs to
(\ref{cubic}) with $z$ replaced by $-z$ shows that the cubic resolvent has no 
negative real roots, and so the quartic (\ref{lenseq8}) does not have
two complex conjugate pairs. Thus, there are always images
present. Furthermore, one can consider the discriminant $\dres$ \cite[
  p. 120]{Br85} of the cubic resolvent to find
\[
\dres=\left(\frac{3B-A^2}{9}\right)^3+\left(\frac{2A^3-9AB+27C}{54}\right)^2\leq 0.
\]
Therefore the cubic resolvent has three positive real roots, and hence
the quartic (\ref{lenseq8}) has four real roots corresponding to four
images which are, in general, distinct. In the limiting case
$\dres=0$, however, double roots occur and we recover the single lens
plane case with two images. This result is in agreement with Erdl \&
Schneider (1993) who have shown, using catastrophe theory, that two
point lenses can produce either four or six images. This has also been
proven by Petters (1995) using Morse theory. It is clear, then, that
the caustic domain giving rise to six images does not occur in our
axisymmetrical case. The limiting behaviour of these four images can
be read off directly from the quartic equation (\ref{lenseq8}). For
$y\rightarrow 0$, there are two images with positive $x$ and two
images with negative $x$ approaching the two Einstein circles. For
$y\rightarrow \infty$ the four images $x_i$ satisfy
\[
x_1\rightarrow y, \ x_2\rightarrow \sqrt{\beta \mo}, \ x_3\rightarrow -\sqrt{\beta \mo}, \ x_4\rightarrow -\frac{\mo}{y}.
\]
Using (\ref{jacob2}) in these cases, we find that $x_2,x_3,x_4$ become
infinitely demagnified so that only $x_1$ with signed magnification
$\mu[x_1]\rightarrow 1$ remains as $y\rightarrow \infty$, as expected.

Furthermore, it turns out that the images obey an invariant of the signed magnification
sum. To see this, we can regard the present case as an example of a more general argument \cite{We07}. 
First, notice that all five possible roots of the lens equation (\ref{lenseq7}) for
$y>0$ are real by the previous discussion such that all of $S\backslash
\bmath{0}$ is a maximal domain. Also, the deflection angle
tends to zero as $y \rightarrow \infty$. Then the lens equation can be complexified such that images correspond to fixed 
points of a complex rational lensing map. This in turn induces a map on complex projective space 
which is holomorphic almost everywhere and, in particular, at the fixed points. Then the holomorphic
 Lefschetz fixed point formula can be applied to yield
\begin{equation}
\sum_{i=1}^5\mu[x_i]=1.
\label{mag}
\end{equation}
In fact, this statement is also true for the four visible images
because $x_5$ is infinitely demagnified as noted above.

\section{Conclusions}

In this article, we have presented an analytical study of multiple
Einstein rings in the weak deflection limit of strong lensing, and now
summarize the three main results. First, it was proven generally that
at most one Einstein ring can occur for one lens plane. A natural and
weak assumption was used here, namely that the normalized surface
density always be smaller than the average surface density. 

Accordingly, we turned to models with lenses in two planes, 
and in which source, observer and both lenses are exactly aligned 
along the optical axis. As a second lens is introduced on the optic 
axis, the existing Einstein ring generally bifurcates. If the more 
distant of the two lenses is also luminous, then it produces a third 
Einstein ring.  In the case of two singular isothermal sphere 
lenses, then the angular radii of the Einstein rings are given by 
equations (\ref{rings}), (\ref{rings2}) and (\ref{rings3}). In the 
case of two point mass lenses, three Einstein rings arise whose 
angular radii are given by (\ref{rings5}), (\ref{rings6}) and 
(\ref{rings7}). These expressions are our second result, the main 
point being that a single source can, in fact, give rise to more 
than one Einstein ring.

We also discussed briefly the image configurations for these two
models if axisymmetry is broken. In the case of the two singular
isothermal lenses, up to two cut circles and four images can arise. In
the case of the two point lenses, we find that there are always four
images. In both instances, the images turn out to possess lensing
invariants -- for example, the magnification invariants
(\ref{isomag}) and (\ref{mag}). This has not been established before
for multi-plane lensing and is our third result.

Finally, we make some critical remarks about possible extensions of
this work. After noting the existence of multiple Einstein rings due
to a single source, the next obvious question is to ask how many there
can be. But counting Einstein rings is clearly more difficult than
counting discrete images. This is because discrete images are
non-degenerate stationary points of the time delay surface, so
theorems of Morse theory can be used, whereas Einstein rings are
degenerate stationary curves and Morse theory does not
apply. Alternatively, the problem of counting Einstein rings can be
set up as a one-dimensional fixed point problem according to equation
(\ref{fix}) within the framework of intersection theory in algebraic
geometry and topology. However, there is a degeneracy problem here,
too, because a non-transverse intersection occurs if the Einstein ring
is degenerate in the sense that, in one lens plane, $\kappa=1$ at this
radius \cite[ p. 234]{Sc99}. Given the large number of free parameters
in the case of multiple lens planes, it will be difficult to ensure
the absence of degenerate Einstein rings.

\section*{Acknowledgments}
MCW gratefully acknowledges financial support by the STFC of the
United Kingdom. JA thanks for the hospitality during the visits to Cambridge.
JA also acknowledges that the Dark Cosmology Centre is funded by the Danish National Research Foundation 
({\it Danmarks Grundforskningsfond}).

\label{lastpage}

\end{document}